# Imaging and structure analysis of ferroelectric domains, domain walls, and vortices by scanning electron diffraction

Ursula Ludacka[1], Jiali He[1], Shuyu Qin[2,3], Manuel Zahn[1,4], Emil Frang Christiansen[5], Kasper A. Hunnestad[1], Zewu Yan[6,7], Edith Bourret[7], István Kézsmárki[4], Antonius T. J. van Helvoort[5], Joshua Agar[2,3], Dennis Meier[1]

Email: dennis.meier@ntnu.no, jca92@drexel.edu

[1]Department of Materials Science and Engineering, NTNU Norwegian University of Science and Technology, Trondheim, Norway
[2]Department of Computer Science and Engineering, Lehigh University, Bethlehem, USA
[3]Department of Mechanical Engineering and Mechanics, Drexel University, Philadelphia, USA
[4]Experimental Physics V, University of Augsburg, Augsburg, Germany
[5]Department of Physics, NTNU Norwegian University of Science and Technology, Trondheim, Norway
[6]Department of Physics, ETH Zurich, Zürich, Switzerland.
[7]Materials Sciences Division, Lawrence Berkeley National Laboratory, Berkeley, USA.

**Direct electron detectors in scanning transmission electron microscopy give unprecedented possibilities for structure analysis at the nanoscale. In electronic and quantum materials, this new capability gives access to, for example, emergent chiral structures and symmetry-breaking distortions that underpin functional properties. Quantifying nanoscale structural features with statistical significance, however, is complicated by the subtleties of dynamic diffraction and coexisting contrast mechanisms, which often results in low signal-to-noise and the superposition of multiple signals that are challenging to deconvolute. Here we apply scanning electron diffraction to explore local polar distortions in the uniaxial ferroelectric $Er(Mn,Ti)O_3$. Using a custom-designed convolutional autoencoder with bespoke regularization, we demonstrate that subtle variations in the scattering signatures of ferroelectric domains, domain walls, and vortex textures can readily be disentangled with statistical significance and separated from extrinsic contributions due to, e.g., variations in specimen thickness or bending. The work demonstrates a pathway to quantitatively measure symmetry-breaking distortions across large areas, mapping structural changes at interfaces and topological structures with nanoscale spatial resolution.**

**Introduction**

High-energy electrons traveling through matter are highly sensitive to the local structure[1], collecting a multitude of information about lattice defects and strain[2], electric and magnetic properties[3], as well as chemical composition and electronic structure[4]. This sensitivity is utilized in transmission electron microscopy (TEM) to study structure-property relations, and there are continuous efforts to increase resolution, enhance imaging speeds, and enable new imaging modalities[5]. A real paradigm shift was triggered by the advent of high dynamic-range direct electron detectors (DED), which no longer rely on converting electrons into photons[6-8]. DEDs enable spatially resolved diffraction imaging, providing new opportunities for high-resolution measurements known as four-dimensional scanning transmission electron microscopy (4D-STEM)[8-10]. A significant advantage of 4D-STEM is the outstanding information density; an image of the dynamically scattered electrons is acquired at every probe position. In turn, advanced analysis tools are required to deconvolute the rich variety of phenomena that contribute to the scattering of the electrons[10-13]. Remarkably, low noise levels on DEDs enable the quantification of weak scattering events (e.g., diffuse scattering due to crystallographic defects[14,15]). The analysis of 4D-STEM data, however, is often challenged by a lack of empirical models that can fully explain the multitude of dynamic scattering processes, as well as varying signal-to-noise ratios. Recently, advances in machine learning have provided a way to disentangle features in multimodal nanoscale spectroscopic imaging with improved statistical significance[16-18]. Through careful design of machine learning architectures and custom regularization strategies, it is now possible to statistically disentangle and interpret structural properties of functional materials with nanoscale spatial resolution from multimodal imaging[19-21].

Here, we apply 4D-STEM to investigate domains, domain walls, and vortex structures in a uniaxial ferroelectric oxide, utilizing the scattering of electrons for simultaneous high-resolution imaging and local structure analysis. Using a convolutional autoencoder (CA) with custom regularization, we statistically disentangle features in the diffraction patterns that correlate with the distinct structural distortions in the ferroelectric domains and domain walls, as well as the domain wall charge state. Based on the specific scattering properties, we can readily gain real space images of ferroelectric domains, domain walls, and their vortex-like meeting points with a resolution limited by the spot size of the focused electron beam (here, 2 nm). Our approach provides a powerful method that combines nanoscale imaging and structural deconvolution – opening a pathway towards improved structure-property correlations, increased fidelity, and automated scientific experiments.

**Results and discussion**

4D-STEM experiments are conducted on a model ferroelectric $Er(Mn_{1-x},Ti_x)O_3$ (x = 0.002), denoted $Er(Mn,Ti)O_3$ in the following. $Er(Mn,Ti)O_3$ is a uniaxial ferroelectric and naturally develops 180° domain walls, where the spontaneous electric polarization $P$ inverts[22-25]. The ferroelectric domain walls have a width comparable to the size of the unit cell[26], and their basic structural[27], electric[26,28], and magnetic properties[29] are well understood, which makes them an ideal model system for exploring local electron scattering events. It is established that the polarization reorientation across the domain walls coincides with a change in the periodic tilt pattern of the $MnO_5$ bipyramids and displacement of Er ions that drive the electric order (i.e., improper ferroelectricity)[24]. The structural changes at domain walls alter the electron scattering processes from the bulk. In turn, this difference is expected to alter scattering intensities encoded in the local electron diffraction patterns obtained in scanning electron diffraction (SED) measurements. There are, however, no good analytical methods to disentangle structural and extrinsic (e.g., thickness- and orientation-related) scattering mechanisms, particularly, in the presence of noise.

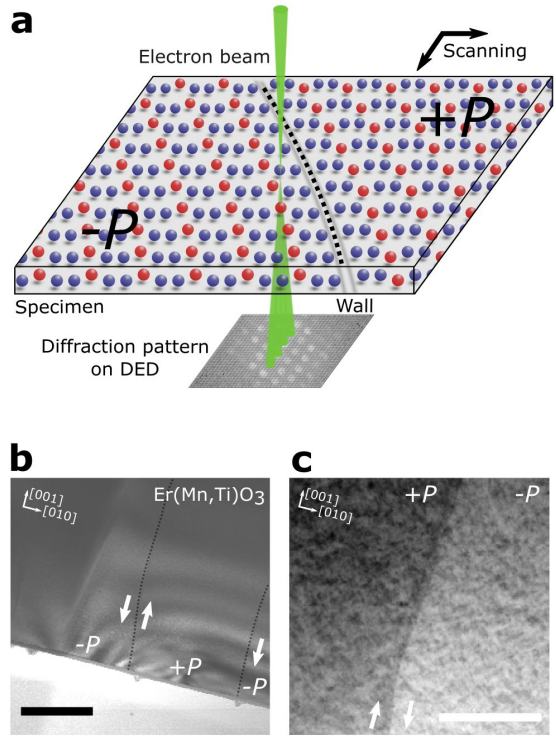

**Figure 1 | Scanning electron diffraction on ferroelectric domains in Er(Mn,Ti)O₃. a,** Schematic of our 4D-STEM approach. The illustration shows how the electron beam (green) is scanned across a domain wall as indicated by the black dashed line, collecting diffraction patterns at a fixed position of the DED. Up-up-down and down-down-up arrangements of red/blue spheres represent the characteristic displacement patterns of Er atoms in the +$P$ and -$P$ domains, respectively. **b,** Overview VDF image showing two ferroelectric domain walls marked by black dashed lines. The bottom part (light gray) is an amorphous carbon layer with Pt markers that were used to cut a lamella from the region of interest. White arrows indicate the polarization direction of the different domains. Scale bar, 250 nm. **c,** High-resolution VDF image recorded at the right domain wall shown in **b**. Scale bar, 100 nm.

The general working principle of SED measurements is illustrated in Figure 1**a**. A focused electron beam is raster-scanned over an electron transparent lamella, extracted from an Er(Mn,Ti)O₃ single crystal using a focused ion beam (FIB, see Methods for details). A diffraction pattern is recorded at each probe position of the scanned area, containing information about the local structure. In addition, integrating and selectively filtering the intensities of the collected individual diffraction patterns allows for calculating virtual real-space images. Figure 1**b** shows such a virtual dark-field (VDF) image. To calculate the VDF, we select and integrate the intensities of the full diffraction patterns as described in ref.[27]. The imaged area contains two ferroelectric 180° domain walls (marked by black dotted lines) that separate +*P* and -*P* domains. The polarization direction within the domains was determined before extracting the lamella from the region of interest based on correlated scanning electron microscopy and piezoresponse force microscopy measurements (not shown). A VDF image with a higher resolution is presented in Figure 1**c** for one of the domain walls, with visible contrast between the two domains. The data in Figure 1**c** is recorded outside the area seen in Figure 1**b** to minimize beam exposure (referred to as data set 1, DS1, in the following).

We begin our discussion of the SED results with a center-of-mass (COM) analysis applied to the complete stack of diffraction patterns in the area presented in Figure 1**c**. The results of the COM analysis are summarized in Figures 2**a** and **b**. In general, the momentum change of the electron probe can be represented by the orientation of a vector in 2D reciprocal space. When interacting with the sample, the direction of the momentum changes, which is used in 4D-STEM COM imaging to determine built-in electric[28] or magnetic[30] fields. To evaluate the COM distribution over the dataset, we plot the COM position of each diffraction pattern as a single spot in reciprocal space. The result gained from the whole dataset is shown in Figure 2**a**, where a substantial redistribution of scattering intensities is observed along the crystallographic [001]-axis (*P* ∥ [001]). We find that the COM shift is sensitive to the local polarization orientation in Er(Mn,Ti)$O_3$, leading to a split in the dispersion line for -*P* (red) and +*P* (blue) domains as seen in Figure 2**b**. Figure 2**b** presents the spatial origin of the two contributions, which coincides with the ferroelectric domain structure resolved in the VDF image in Figure 1**c**.

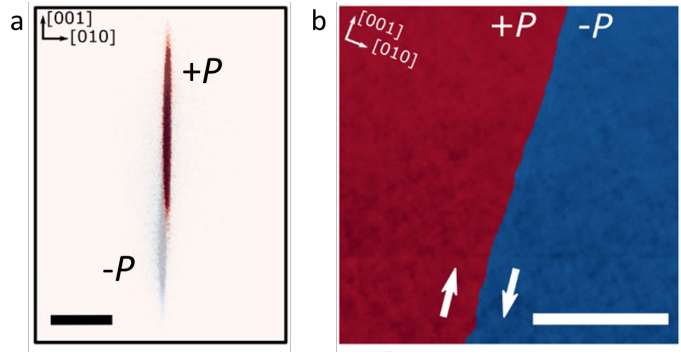

**Figure 2 | Domain-dependent scattering of electrons. a,** The center-of-mass (COM) analysis of every diffraction pattern in DS1 shows a substantial shift with respect to the geometric center in the upwards (downwards) direction along the crystallographic [001]-axis for -$P$ (+$P$) domains. Scale bar, 0.1 Å$^{-1}$. **b**, COM analysis of the diffraction patterns associated with -$P$ (red) and +$P$ (blue) domains. Scale bar, 100 nm.

To analyze the domain-dependent scattering in more detail, we deploy a custom CA. The autoencoder consists of different blocks as illustrated in Figure 3**a-c**. The CA takes the input diffraction patterns and learns a low-dimensional statistical representation of the image through a series of convolutional and residual blocks. In each residual block, a max pooling (MaxPool) layer reduces the dimensionality of the image. Once the dimensionality of the image is sufficiently reduced, the two-dimensional image is flattened into a feature vector. This penultimate bottleneck layer is further compressed to a low-dimensional latent space, where statistical characteristics of the structure are disentangled using a scheduled custom regularizer. The learned latent representation is reshaped into a 2D image and decoded in the decoder using a series of upsampling residual blocks until the image is reconstructed to its original resolution. The model is trained using momentum-based stochastic gradient descent (ADAM) to minimize the mean squared reconstruction error of the diffraction images and regularization constraints added to the loss function.

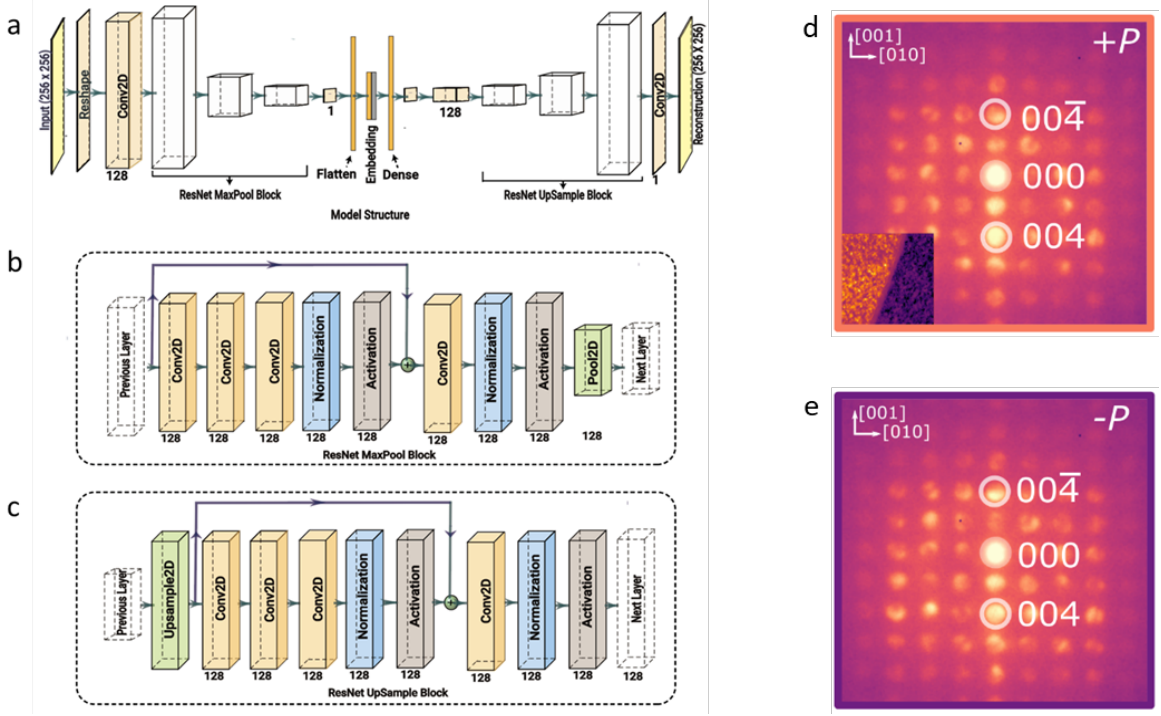

**Figure 3 | Structure of the custom CA. a,** Main structure, consisting of encoder (from input to flatten layer), embedding and decoder (from dense layer to reconstruction). The encoder reduces the dimension of each input image by going from 256 x 256 pixels to 8 x 8 pixels and via a dense layer down to the embedding. The embedding controls the number of channels to generate individual domains and domain walls in real space. The decoder recreates the vector from the embedding to the input image size. **b,** Detailed structure of the ResNet MaxPool Block. The block consists of four convolutional layers, two layer-normalization layers, two ReLU activation layers, and one 2D MaxPool layer with shortcut. **c,** Detailed structure of the ResNet UpSample Block. The block contains one 2D upsample layer, four convolutional layers, two-layer normalization layers, and two ReLU activation layers with shortcut. **d,** Averaged diffraction pattern of a +P domain in dataset DS1, corresponding to the left domain (orange) seen in the CA embedding in the inset. **e,** Averaged diffraction pattern of the -P domain (purple) in the CA embedding in the inset to **d**.

We impose various constraints on the embedding layer to encourage interpretable disentanglement of ferroelectric domains in the latent space. First, we add a rectified linear activation (ReLu) to ensure the activations are non-negative. All neural networks have a loss function based on the mean squared reconstruction error $MSE(y, \hat{y}) = \frac{1}{D}\sum_{i=1}^{D}(y_i - \hat{y}_i)^2$, where $y$ and $\hat{y}$ denote the $D$-dimensional output and input of the neural network ($D = 256^2 = 65,536$), respectively. To impose sparsity (a limited number of activated channels), an additional activity regularization is introduced $L_1(a) = \sum_{i=1}^{d}|a_i|$, leading to a total loss function

$$L = MSE(y, \hat{y}) + \lambda_{act}L_1(a); \qquad (1)$$

here, $d$ is the is the dimensionality of the embedding layer, $a_i$ are the activations in the embedding layer, and $\lambda_{act}$ is a hyperparameter. This has the effect of trying to drive most activations to zero while only those essential to the learning process are non-zero. As the degree of sparsity required is dataset-dependent, regularization scheduling is used to tune $\lambda_{act}$ to achieve an interpretable degree of disentanglement.

To demonstrate the efficiency of the CA, we analyze 4D-STEM data from the region with two ferroelectric domains seen in Figure 1**c** (DS1). The model is trained with an overcomplete embedding layer of size 32. Following training, the number of active channels is reduced to 9 (see Supplementary Note 1 and Figure S1). Most of the embeddings disentangle bias in the imaging mode associated with the scan geometry, varying specimen thickness and orientation variations due to specimen bending; additionally, features associated with the domain wall are disentangled, which we will discuss later. One channel shows a sharp contrast between the 180° domains, indicating a significant contrast mechanism (inset to Figure 3**d**). This map represents the activations of one neuron and, hence, is a weighting map for a specific characteristic in the diffraction pattern. To elucidate the nature of the contrast mechanism, we traverse the neural network latent. We show the generated diffraction patterns from the latent space encompassing the +*P* and -*P* domains in Figure 3**d,e**.

The CA analysis reveals variations between the two domain states in the scanned area for the strongest reflections along the [001]-axis, that is, the 004 and $00\bar{4}$ reflections (note that intensity distributions vary with sample thickness). A substantial advantage of the CA-based approach compared to, e.g., signal decomposition via unsupervised non-negative matrix factorization, is that it does not create artificial components that resemble diffraction patterns. Instead, the CA rates each diffraction pattern according to the scattering features in the embedding channels. Thus, by selecting and averaging diffraction patterns within a specific activation range within a certain channel, one can readily use this approach as a virtual aperture in reciprocal space using multiple areas of the pattern to correlate structural features identified statistically to scattering properties.

To demonstrate that the diffraction patterns in Figure 3**d,e** are indeed specific to the local polarization orientation and connect them to the atomic-scale structure of Er(Mn,Ti)O$_3$, we simulate the diffraction patterns expected for +$P$ and -$P$ domains using a Python multislice code[31]. As one example, Figure 4**a** displays the unit cell structure of a +$P$ domain, which is reflected by the up-up-down pattern formed by the Er atoms[32]. The corresponding simulated diffraction pattern is presented in Figure 4**b**, considering a sample thickness of 75 nm. Figure 4**b** shows an asymmetry in the 004 and 00$\bar{4}$ reflections, consistent with the diffraction data in Figure 3**d**. For a more systematic comparison of the experimental and simulated diffraction patterns, we calculate the normalized cross-correlation $\Delta$(+$P$, -$P$) between the patterns of the two domains, as shown in Figure 4**c** (simulated) and Figure 4**d** (experimental). In both cases, the variational maps show the highest intensities wherever the two compared patterns exhibit the strongest variations. As expected, those arise primarily in the 004 (00$\bar{4}$) and, less pronounced, in the 002 (00$\bar{2}$) reflections. This observation further corroborates the CA-based analysis of the SED data, linking the changes in the diffraction pattern intensities to the atomic displacements and the resulting polarization direction.

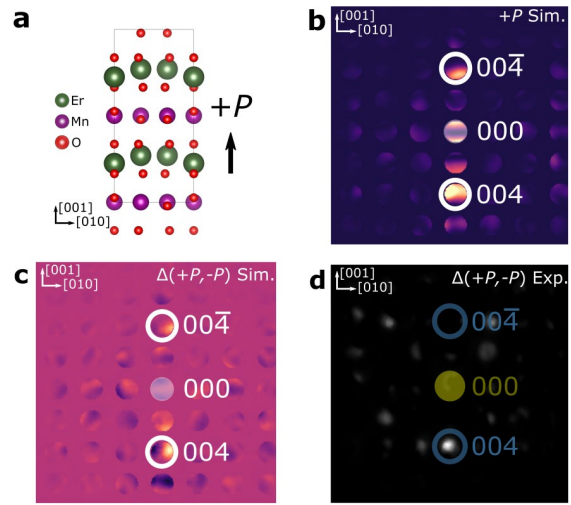

**Figure 4 | Comparison of measured and simulated SED diffraction patterns. a,** Illustration of the atomic structure in +$P$ domains, showing the characteristic up-up-down displacement pattern of Er atoms. The crystallographic [001] and [010] axes are indicated by the inserted coordinate systems. **b,** Simulated diffraction pattern for the structure in **a**. The 004 and $00\bar{4}$ reflections are marked by white circles. **c,** Normalized cross-correlation between simulated (**d,** experimental, DS1) diffraction patterns of -$P$ and +$P$ domains, $\Delta(+P, -P)$, showing that the highest variation occurs for the 004-reflection.

After demonstrating that our approach is sensitive to the polar distortions in Er(Mn,Ti)$O_3$, and that it can extract domains, we discuss local variations in the diffraction pattern intensities that originate from finer structural changes. Figure **5a** displays the same embedding map as seen in the inset to Figure **3d**, showing two ferroelectric domains with opposite polarization orientation. A second embedding map is shown in Figure **5b**, indicating scattering variations at the position of the domain wall (see also Supplementary Note 1 and Figure S1). The latter reflects the broader applicability of the CA beyond domain-related investigations. To explore the possibility to investigate local structure variations also at domain walls, we conduct additional measurements on a sample with multiple walls that meet in a characteristic six-fold meeting point, leading to a structural vortex pattern[23,26,29,33] as presented in Figure **5c-f** (referred to as DS2). It is established that such vortices promote the stabilization of different types of walls[34], which allows for testing the feasibility of our 4D-STEM approach for structure analysis of ferroelectric domain walls with varying physical properties.

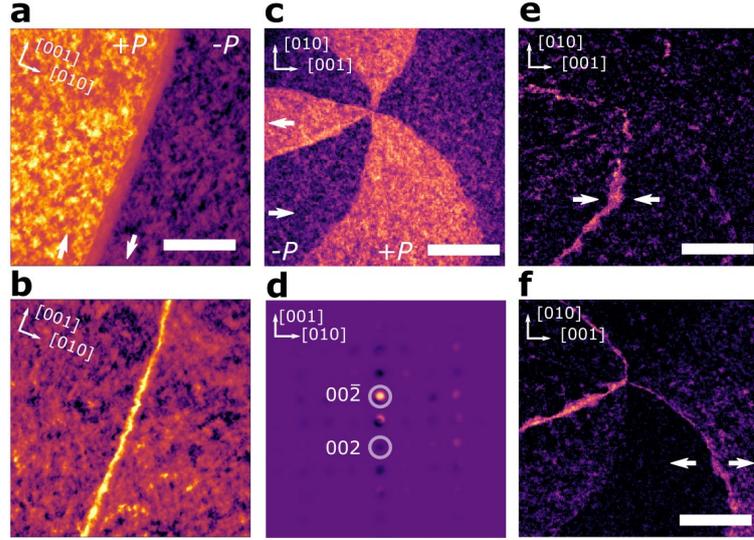

**Figure 5 | Domains and domain walls extracted via the CA. a,** Embedding map showing ferroelectric ±$P$ domains (DS1). Scale bar, 75 nm. Polarization directions are given by white arrows (same as inset to Figure 2**c**). **b,** Embedding map revealing the domain wall that separates the domains in **a**. **c,** Embedding map from a second sample (DS2). Scale bar, 90 nm. **d,** Difference in diffraction patterns between +$P$ and -$P$ domains in **c**. **e, f,** Two embedding maps of the CA, separating head-to-head (**e**) and tail-to-tail (**f**) domain walls that belong to the vortex in **c**.

As the statistics of the domain walls are different than within the domains, a uniform sparsity metric cannot disentangle these features well. Thus, to improve the performance of our model, we add two additional regularization parameters to the loss function that encourage sparsity and disentanglement. First, we add a contrastive similarity regularization of the embedding, $L_{sim}$, to the loss function. This regularization term computes the cosine similarity between each of the non-zero vectors $a_i$ and $a_j$ within a batch of embedding vectors, where $N_{batch}$ is the batch size, and $\lambda_{sim}$ is a hyperparameter that sets the relative contribution to the loss function.

$$L_{sim} = \frac{\lambda_{sim}}{2N_{batch}} \sum_{i=1}^{N_{batch}} \left( \left[ \sum_{j=1}^{N_{batch}} \frac{a_i \cdot a_j}{||a_i|| \cdot ||a_j||} \right] - 1 \right) .$$

Since the activations are non-negative, the cosine similarity is bounded between [0,1], where 0 defines orthogonal vectors, and 1 defines parallel vectors. We subtract 1, so that similar and sparse vectors have no contribution to the loss function, whereas dissimilarity of non-sparse vectors decreases the loss and, thus, is encouraged.

Secondly, we add an activation divergence regularization, $L_{div}$, to the loss function, where $a_{i,j}$, $a_{i,k}$ are components of the $i^{th}$ vector within a batch of latent embeddings. The magnitude of this contribution is regulated using the hyperparameter $\lambda_{div}$:

$$L_{div} = \frac{\lambda_{div}}{2N_{batch}} \sum_{i=1}^{N_{batch}} \left( \sum_{j=1}^{d} \sum_{k=1}^{d} |a_{i,j} - a_{i,k}| \right).$$

This term has the effect of enforcing that each embedding vector is sparse, having a dominate component that is easy to interpret. We use the hyperparameter $\lambda_{div}$ to ensure that the magnitude of this contribution is significantly less than the reconstruction error. When applying these custom regularization strategies, the resulting activations disentangle more nuanced features in the domain structure.

The model readily disentangles the +$P$ and -$P$ domain states as presented in Figure 5c, revealing a six-fold meeting point of alternating ±$P$ domains. The difference pattern between the two domain states can be determined using the CA as a generator. To do so, we calculate the mean pattern of the upper 5% quantile of the +$P$ (orange) and -$P$ (purple) domains in Figure 5c, which leads us to Figure 5d (corresponding color histograms are shown in Supplementary Figure S2). Consistent with Figure 4, pronounced intensity variations between +$P$ and -$P$ domains are observed for the 004 ($00\bar{4}$) and 002 ($00\bar{2}$) reflections. In contrast to the data collected on the first sample (Figure 4), however, Figure 5d reveals a stronger variation in the 002 ($00\bar{2}$) reflections, which we attribute to a difference in sample thickness.

Interestingly, the neural network produces different embedding maps for the domain walls in Figure 5c, indicating a difference in their scattering behavior. Specifically, we disentangle statistical features that reveal the existence of two sets of domain walls as shown in Figure 5e,f, respectively (additional embeddings are shown in Supplementary Figure S2). Based on the polarization direction in the adjacent domains, we can identify the two sets of domain walls as negatively charged tail-to-tail walls (Figure 5e) and positively charged head-to-head walls (Figure 5f). This separation regarding the polarization configuration is remarkable as it reflects that our approach is sensitive to both the crystallographic structure of the domain walls and their electronic charge state as defined by the domain wall bound charge[33].

In summary, our work demonstrates a new pathway for imaging and characterizing ferroelectric materials at the nanoscale. By applying a custom-designed CA to SED data gained on the model system $Er(Mn,Ti)O_3$, we have shown that different scattering signatures can be separated within the same experiment. The latter includes ferroelectric domains, domain walls, and emergent vortex structures, as well as extrinsic features (e.g., bending and thickness variations), giving access to both the local structure and electrostatics. The findings can readily be expanded to other systems to localize, identify, and correlate weak scattering signatures to structural variations based on SED. By building a CA with custom regularization to promote disentanglement, subtle spectroscopic signatures of structural distortions can be statistically unraveled with nanoscale spatial precision. This approach is promising to automate and accelerate the unbiased discovery of defects, secondary phases, boundaries, and other structural distortions that underpin functional materials. Furthermore, it opens the possibility to expand the design of experiments to larger imaging sizes, higher frame rates, and more broadly into automated experimentation and, eventually, controls.

# Acknowledgements

U.L. thanks Muhammad Z. Khalid for help with the visualization of the crystal structure. D.M., U.L., and J.H. acknowledge funding from the European Research Council (ERC) under the European Union's Horizon 2020 Research and Innovation Program (Grant Agreement No. 863691). D.M. thanks NTNU for support through the Onsager Fellowship Program and the Outstanding Academic Fellow Program. The Research Council of Norway is acknowledged for the support to the Norwegian Micro- and Nano-Fabrication Facility, Nor-Fab, project number 295864 and the Norwegian Center for Transmission Electron Microscopy, NORTEM (197405). S.Q. acknowledges support from the National Science Foundation under grant TRIPODS + X:RES-1839234 and DOE Data Reduction for Science award Real-Time Data Reduction Codesign at the Extreme Edge for Science. J.C.A. acknowledges support from the Army/ARL via the Collaborative for Hierarchical Agile and Responsive Materials (CHARM) under cooperative agreement W911NF-19-2-0119, and National Science Foundation under grant OAC:DMR:CSSI - 2246463. M.Z. acknowledges funding from the Studienstiftung des Deutschen Volkes via a Doctoral Grant and the State of Bavaria via a Marianne-Plehn scholarship.

# Author contributions

U.L. performed the 4D-STEM measurements supervised by D.M and with support from E.F.C. and A.T.v.H; J.H. conducted SPM and SEM measurements and prepared the lamellas for the SED experiments. S.Q. designed, trained, and analyzed the convolutional autoencoder under supervision by J.C.A., with support from U.L. and M.Z. (supervised by D.M. and I.K.). K.A.H. simulated the diffraction pattern assisted by U.L., Z.Y. and E.B. provided the single crystals used in this work. D.M. initiated and coordinated the project. U.L., M.Z., J.C.A., and

D.M. wrote the manuscript. All authors discussed the results and contributed to the final version of the manuscript.

## Methods

**Specimen preparation.** The TEM lamellas were prepared from an $Er(Mn,Ti)O_3$ single crystal using a Thermo Fisher Scientific G4 UX DualBeam FIB. Ion milling was conducted at 90 pA and a final electron beam polishing step with a voltage of 2 kV and a current of 0.2 nA to minimize the beam damage. The positions of domain walls were marked by deposited Carbon and Platinum. The resulting lamellas had a thickness of maximum 75 nm, which was determined during the FIB preparation via the specimen contrast.

**Diffraction data acquisition.** The diffraction experiments were conducted on a Jeol 2100F TEM at 200 kV and the scans were controlled via the Nanomegas P1000 scan engine. For acquiring the diffraction patterns, we used a Merlin 1S DED from Quantum detectors operated with a lower threshold of 40 kV and with no limit on the upper threshold. The electron beam is focused to a probe with a diameter of 2 nm and a convergence angle of 9 mrad. The total scan grid consisted of 256 x 256 probe positions (with a step size of 1.4 nm) with a probe dwell time of 50 ms at each beam position.

**Convolutional Autoencoder (CA).** Data from 4D-STEM was analyzed using a CA built in Pytorch[35]. Prior to training, the log of the raw 4D-STEM data was used to obtain less non-linear images. The number of learnable parameters is 4,700,770. The CA consists of three parts, an encoder, an embedding layer and a decoder. The encoder consists of three ResNet Blocks with different feature size, a convolutional layer with one filter and a flatten layer. Each ResNet Block consists of a Residual Convolutional Block and an Identity Block. Each Residual Convolutional Block has three sequence convolutional layers with 128 filters, connected with a normalization layer and a Rectified Linear Unit (ReLU) activation layer. There is a skip connection between the input and output of the block, which can maintain the information of the input image after image processing. Each Identity Block has a convolutional layer with 128 filters, connected with a normalization layer and a ReLU activation layer. There is a 2D Max Pooling layer after each Resnet Block for image size dimensionality reduction. The image sizes to each ResNet Block in the encoder are (256 x 256), (64 x 64), (16 x 16). The embedding consists of a linear layer and a ReLU activation layer. The decoder consists of a linear layer, a convolutional layer with 128 filters, three ResNet Blocks and a convolutional layer with 1 filter. There is an upsampling layer before each ResNet Block to recreate the input image. A loss function based on the mean square reconstruction error (MSE) between the input and generated image is used. The image sizes to each ResNet Block in the decoder are (8 x 8), (16 x 16), (64 x 64). The loss function has additional $L_1$ activity regularization of the embedding. When generating domain walls in Figure 5e and 5f, we also include contrastive similarity regularization and activate divergence regularization to make the output embedding sparse and unique.

The models were trained on a server with 4x A100 GPUs. To generate the domain in Figure 5a and the domain wall in Figure 5b, we set the coefficient $\lambda_{act} = 1 \cdot 10^{-5}$ and trained the model using optimization ADAM[36] (learning rate of $3 \cdot 10^{-5}$) for 377 epochs. To generate the vortex-like domain pattern (Figure 5c), we set the coefficient $\lambda_{act} = 1 \cdot 10^{-5}$ and trained the model for 225 epochs using optimization ADAM (learning rate of $3 \cdot 10^{-5}$), then raised $\lambda_{act}$ to $5 \cdot 10^{-4}$ and trained the model for another 60 epochs using learning rate cycling (increasing from $3 \cdot 10^{-5}$ to $5 \cdot 10^{-5}$ in 15 epochs, then decreasing from $5 \cdot 10^{-5}$ to $3 \cdot 10^{-5}$ in next 15 epochs). To generate the corresponding domain walls in Figure 5e and 5f, besides $L_1$ regularization with coefficient $\lambda_{act} = 5 \cdot 10^{-3}$ in the loss function, we also included contrastive similarity regularization with coefficient $\lambda_{sim} = 5 \cdot 10^{-5}$ and activity divergence regularization with coefficient $\lambda_{div} = 2 \cdot 10^{-4}$ to make the output embedding sparse and unique. We trained the model for 18 epochs using optimization ADAM (learning rate of $3 \cdot 10^{-5}$). Following training, the output from the embedding layer was extracted. This represents a compact representation of the important features in the sample domain. To visualize the change in the diffraction pattern that is encoded by a single channel, the difference between the mean pattern of all diffraction pattern with 5% highest and lowest activation at the channel of interest was calculated. This was used to create the projections in Figures 3**c, d**, Figure 4**d,** Figure 5**d** and the third row of Supplementary Figure S2. Full details are available in the reproducible source code.

### Data Availability:

All data and reproducible code are made openly available under the BSD-2 License. The raw 4D-STEM data is published on Zenodo via DOI 10.5281/zenodo.7837986. The source code is built as part of the M3-Learning[37]. Some of the core modules can be installed using the command `pip install m3_learning`. The full release is available on Zenodo[38] via DOI 10.5281/zenodo.7844268. To improve accessibility, a Jupyter Notebook is available.

# Supplemental Information

# Imaging and structure analysis of ferroelectric domains, domain walls, and vortices by scanning electron diffraction

Ursula Ludacka[1], Jiali He[1], Shuyu Qin[2,3], Manuel Zahn[1,4], Emil Frang Christiansen[5], Kasper A. Hunnestad[1], Zewu Yan[6,7], Edith Bourret[7], István Kézsmárki[4], Antonius T. J. van Helvoort[5], Joshua Agar[2,3], Dennis Meier[1]

Email: dennis.meier@ntnu.no, jca92@drexel.edu

[1]Department of Materials Science and Engineering, NTNU Norwegian University of Science and Technology, Trondheim, Norway

[2]Department of Computer Science and Engineering, Lehigh University, Bethlehem, USA

[3]Department of Mechanical Engineering and Mechanics, Drexel University, Philadelphia, USA

[4]Experimental Physics V, University of Augsburg, Augsburg, Germany

[5]Department of Physics, NTNU Norwegian University of Science and Technology, Trondheim, Norway

[6]Department of Physics, ETH Zurich, Zürich, Switzerland.

[7]Materials Sciences Division, Lawrence Berkeley National Laboratory, Berkeley, USA.

**Supplementary Note 1: Analysis of convolutional autoencoder data**

The input of the convolutional autoencoder (CA) was the raw dataset of $256^2 = 65{,}536$ diffraction patterns, which delivered active embedding channels of specimen scattering characteristics. The common scattering features of specific areas are visualized by a color histogram with 200 bins. The top rows in Supplementary Figure S1 and S2 show all active embeddings for both datasets covered in the main text; the bottom row in Supplementary Figure S1 presents the corresponding histograms of embedding activations. The second row of Supplementary Figure S2 shows the averaged diffraction pattern of a selected embedding feature with low embedding activation. The third row in Supplementary Figure S2 shows the difference pattern between high and low embedding activation of the corresponding channel. We obtained the resulting patterns of the scattering features (originating from domain and domain wall) by averaging diffraction patterns around the characteristic embedding activations. In the case of the embedding showing domain contrast, the histograms (bottom right) show two significant peaks. Until now, it is important to know what scattering characteristic a specific specimen feature exhibits, like features originating from specimen mistilt (embedding channel 2 in Supplementary Figure S2) or thickness variations (embedding channel 5 in Supplementary Figure S2). These can then be correlated with the correct embedding channel and simulations, such as in the case of the differences between domains.

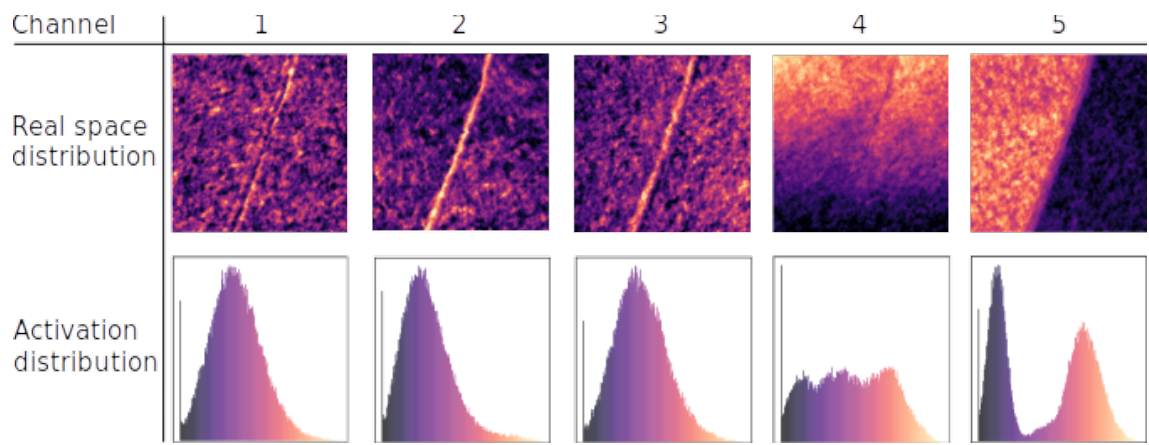

**Figure S1 | Active embedding channels – neutral domain wall (DS1).** Resulting channels, showing real space rating maps of common electron scattering features; histograms of the embedding intensities are presented in the bottom row.

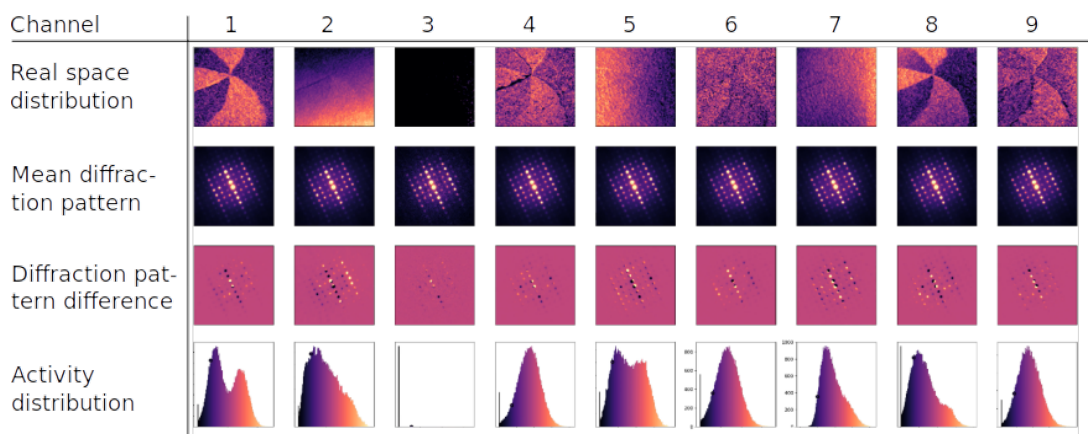

**Figure S2 | Active embedding channels – vortex structure (DS2).** Active resulting embeddings for dataset DS2 with low $L_1$ activity regularization. The mean diffraction patterns represent an average over 5% of the diffraction patterns, including those with the lowest activation within the embedding channel of interest.